\documentclass{article}

\usepackage{PRIMEarxiv}

\usepackage[utf8]{inputenc} 
\usepackage[T1]{fontenc}    
\usepackage{hyperref}       
\usepackage{url}            
\usepackage{booktabs}       
\usepackage{amsfonts}       
\usepackage{nicefrac}       
\usepackage{microtype}      
\usepackage{lipsum}
\usepackage{fancyhdr}       
\usepackage{graphicx}       
\graphicspath{{media/}}     
\usepackage{textcomp}

\usepackage{listings}
\usepackage{amsthm}
\usepackage{amssymb}
\usepackage{subfigure}
\usepackage{multicol}
\usepackage{ctable}
\usepackage{float}
\usepackage{times}
\usepackage{pifont}
\usepackage{upgreek}
\usepackage{algorithmicx}
\usepackage{algorithm}
\usepackage{algpseudocode}
\usepackage{multirow}
\usepackage{caption}

\title{SafeSoftDR: A Library to Enable Software-based Diverse Redundancy for Safety-Critical Tasks
}

\author{
  Fabio Mazzocchetti$^\dagger$, Sergi Alcaide$^{\dagger,\ddagger}$, Francisco Bas$^{\dagger,\ddagger}$, Pedro Benedicte$^\dagger$,\\
  \textbf{Guillem Cabo$^\dagger$, Feng Chang$^\dagger$, Francisco Fuentes$^\dagger$, Jaume Abella$^\dagger$}\\
  $^\dagger$Barcelona Supercomputing Center (BSC), Spain\\
  $^\ddagger$Universitat Politecnica de Catalunya (UPC), Spain \\
}

\begin{document}
\maketitle

\begin{abstract}
Applications with safety requirements have become ubiquitous nowadays and can be found in edge devices of all kinds. However, microcontrollers in those devices, despite offering moderate performance by implementing multicores and cache hierarchies, may fail to offer adequate support to implement some safety measures needed for the highest integrity levels, such as lockstepped execution to avoid so-called \emph{common cause failures} (i.e., a fault affecting redundant components causing the same error in all of them). To respond to this limitation, an approach based on a software monitor enforcing some sort of software-based lockstepped execution across cores has been proposed recently in \cite{SergiDFT}, providing a proof of concept. 
This paper presents SafeSoftDR, a library providing a standard interface to deploy software-based lockstepped execution across non-natively lockstepped cores relieving end-users from having to manage the burden to create redundant processes, copying input/output data, and performing result comparison.
Our library has been tested on x86-based Linux and is currently being integrated on top of an open-source RISC-V platform targeting safety-related applications, hence offering a convenient environment for safety-critical applications.
\end{abstract}

\section{Introduction}

Applications with safety requirements have become increasingly ubiquitous nowadays and can be found in cars, satellites, industrial applications, smart cities, smartphones, etc. These systems are often equipped with multicore processors with varying degrees of performance, but it is not uncommon that they implement scarce, or at least limited, support for safety. This is, for instance, the case for ARM Cortex-A and Cortex-R families and NXP P/T/LS families. Even processors targeting the space domain, such as Gaisler's LEONx families, while providing abundant support for reliability, do not implement some features such as dual-core lockstep (DCLS). Such type of support is typically available in other types of products, such as Infineon AURIX processors for the automotive domain. Still, lack the flexibility to use lockstepped cores independently to run different applications when safety requirements are low (or null). 

DCLS is generally needed by high-integrity applications requiring support to make residual the risk of a single fault leading the system to a failure. While redundancy is effective against many fault types, redundancy on its own is not enough if a fault affects redundant components similarly. For instance, two redundant cores executing the same program synchronized could experience a fault in their common clock input signal or power supply, which could lead both of them to the same erroneous output, which would not be detected employing comparison. DCLS imposes staggered execution across redundant cores (e.g., 2-3 cycles of delay for one of the cores w.r.t. the other). Upon a common fault, both cores have a different internal state and experience different errors that can be detected through comparison. However, as discussed in \cite{SergiDFT}, such an approach is generally too expensive and inflexible since redundant cores cannot be used independently, hence wasting half of the potential performance of the platform even if DCLS is not needed.

Recently, we have proposed a solution based on a software monitor able to enforce staggered execution and hence diversity across redundant user processes~\cite{SergiDFT}. However, such a solution has only been prototyped to prove the feasibility of the approach, but a standard interface has not been offered, relieving end-users from the burden of having to replicate input data, create redundant processes and compare results. 

This paper presents SafeSoftDR, a library implementing diverse redundancy support with software-only means, hence compatible with any multicore lacking native DCLS, that relieves end users from the burden to manage the process. As shown in this paper, such library is feasible, provides a standard interface for end-users, and is successfully deployed on a Linux-based Intel multicore -- it was already proven compatible with ARM multicores in \cite{SergiDFT}. Such library is currently being ported to RISC-V and, whenever its validation is complete, it will be offered as an open-source component with a permissive license in \cite{CAOSgit} along with a number of already public safety-related components.
\section{State-of-the-Art}

Hardware support for safety has been recently reviewed in a survey \cite{SurveyMulticores}. 
Some solutions in the literature have considered how to achieve redundancy across different cores, but without providing diversity~\cite{Mukherjee2002,Gomaa2003,dynamic_coupled_cores}, as needed to avoid common cause failures.
The most relevant works for our paper are those providing diverse redundancy for cores. Among those, we identify CPUs providing native DCLS support, such as, for instance, the Infineon AURIX processor family~\cite{infineon_aurix} and ST Microelectronics SPC56XL70~\cite{STlockstep}. Also, some works have shown how to achieve some form of flexible DCLS support with hardware means. In the case of CPUs, this is achieved with a hardware monitor enforcing staggering (SafeDE)~\cite{SafeDE} or a hardware monitor measuring diversity (SafeDM)~\cite{SafeDM}. In the case of GPUs, this is achieved with the integration of an ad-hoc scheduler~\cite{SergiDATE}. However, those solutions need explicit software support and cannot be used in Commercial-off-the-Shelf (COTS) multicores without any such support. 
To cover this gap, our recent work has provided a proof of concept of the feasibility of a software-only solution to enforce staggered execution across cores running a given application redundantly~\cite{SergiDFT}. While such a solution is naturally less efficient than those building on hardware support, it is compatible with COTS platforms.

In this paper, we generalize the solution in \cite{SergiDFT} by building a portable library implementing software-based DCLS relieving end-users from the burden of managing input data replication, redundant process creation and result comparison.

\section{SafeSoftDR Library}

\textbf{The concept}. SafeSoftDR implements a software monitor able to keep a given staggering between two redundant processes. The staggering is measured in the number of instructions, and it is a parameter for SafeSoftDR. The monitor checks the staggering across the head and trail processes at a given frequency -- also a SafeSoftDR parameter -- and, if the current staggering is below the corresponding threshold, the monitor stops the trail process. If the staggering is above the threshold and the trail process is stopped, the monitor activates the trail process so that it can resume its execution.

As discussed in \cite{SergiDFT}, the staggering between the head and tail processes must be sufficiently large so that, if the head process gets completely stalled and the trail one runs at full speed right after the monitor checks that the staggering is enough, the monitor must be in time to stop the trail process in the next checking interval. Such staggering is completely platform dependent since it is determined by the peak performance of the cores in the platform, and the overheads imposed by the operating system to retrieve instruction counts from the head and trail cores to the core where the monitor runs, as well as to stop the trail process if needed. In general, such threshold needs to be determined empirically, but recommendations in \cite{SergiDFT} provide guidance on how to do it.

\begin{figure}[h!]
\centering
\begin{tabular}{|l|}
  \hline
\begin{lstlisting}[language=C++,basicstyle=\ttfamily\scriptsize]
void matrix_multiply_wrapper(void * argv_input[], void * argv_output[] ){
    int *matA = (int * )argv_input[0];
    int *matB = (int * )argv_input[1];
    int *matC = (int * )argv_output[0];

    int rows = *(int *)argv_input[2];
    int cols = *(int *)argv_input[3];

    matrix_mutilply(matA, matB, matC, rows, cols);
}
\end{lstlisting} \\
  \hline
\end{tabular}
  \caption{Matrix multiplication wrapper}
  \label{fig:wrap}
\end{figure}

\textbf{The interface}. To allow the SafeSoftDR to generate redundant processes replicating input and output data and comparing results, it needs to receive information in a particular format, which requires end-users to create a wrapper for their software to be protected. In particular, such wrapper needs to be invoked with a vector with pointers to the input data and another with pointers to the output data. This is, for instance, illustrated in Figure~\ref{fig:wrap} for the example of matrix multiplication. The wrapper, \texttt{matrix\_multiply\_wrapper}, calls the matrix multiplication function (\texttt{matrix\_multiply}) unfolding input and output data vectors, as shown in the figure.

\begin{figure}[!ht]
\centering
\begin{tabular}{|l|}
  \hline
\begin{lstlisting}[language=C++,basicstyle=\ttfamily\scriptsize]
void (*ptr)( void *[], void *[]) = &matrix_multiply_wrapper;
void * argv_input[] = { (void *)matA, (void *)matB, (void *)&rows, (void *)&cols, NULL};
void * argv_output[] = { (void *)matC, NULL};
int * input_size[] = {(int *)matAbytes, (int *)matBbytes, (int *)&rowsbytes, (int *)&colsbytes, NULL};}
int * output_size[] = { (int *)matCbytes, NULL};

bool pass_flag = protect_default(ptr, argv_input, input_size, argv_output, output_size) ;
\end{lstlisting} \\
  \hline
\end{tabular}
  \caption{Call to the monitor passing as argument the matrix multiplication wrapper}
  \label{fig:call}
\end{figure}

The SafeSoftDR library is used as illustrated in Figure~\ref{fig:call} for the example of the matrix multiplication. In particular, the monitor, \texttt{protect\_default}, needs receiving the following five parameters:
\begin{itemize}
\item A pointer to the application wrapper, \texttt{matrix\_multiply\_wrapper} in the example.
\item The vector with the pointers to the input data to use, \texttt{argv\_input}.
\item A vector, \texttt{input\_size}, with the size of each input data item in \texttt{argv\_input}.
\item The vector with the pointers to the output data to produce, \texttt{argv\_output}.
\item A vector, \texttt{output\_size}, with the size of each output data item in \texttt{argv\_input}.
\end{itemize}

Note that if a particular data item is to be used as input/output data, this needs to be managed accordingly by the application wrapper. In this case, as in any other case, SafeSoftDR will create independent copies of the input and output data for each of the redundant wrapper invocations. The wrapper should use pointers as needed to operate on the copy of the data in the output vector. This typically will imply dismissing the data in the input vector and passing the pointer to the output vector to the application as both input and output data.

\begin{figure}[!h]
\centering
  \includegraphics[width=0.35\textwidth]{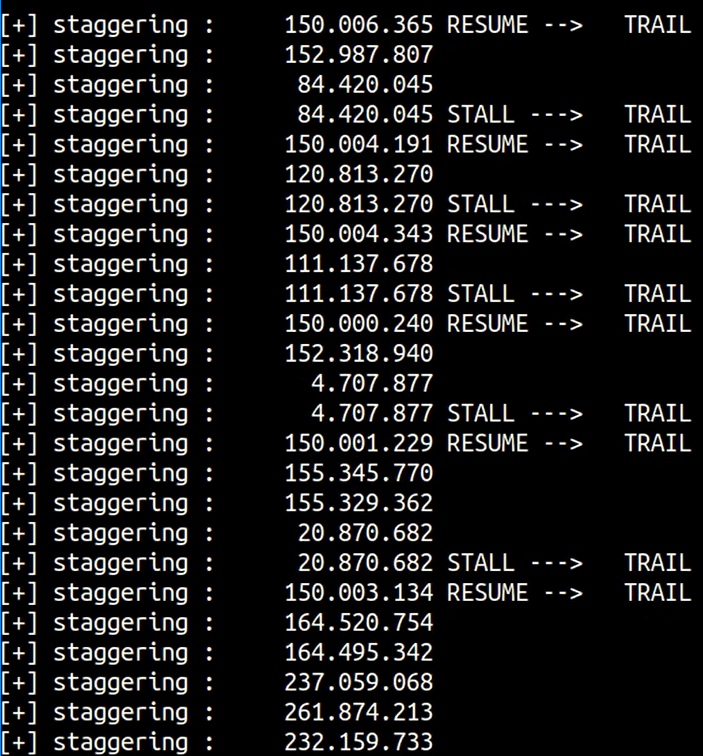} 
  \caption{Example of use of SafeSoftDR}
  \label{fig:result}
\end{figure}

\textbf{An illustrative example}. To showcase SafeSoftDR, we have created a demo of this technology in which the matrix multiplication described before is executed with diverse redundancy. In particular, the matrix multiplication has been set sufficiently large so that it takes more than 1 minute to run for the sake of the demo. We have deployed SafeSoftDR on an {\color{black}Ubuntu 18.04 release running on an Intel\textregistered ~Core\texttrademark ~i7-5600U CPU at 2.6GHz with 16GB of DRAM}. For demo purposes, we print the actual staggering (in number of instructions) every time the monitor is executed in its own core. However, since printing on screen is a slow process subject also to some execution time variability on top of a regular Linux operating system, we have set the staggering threshold to be large enough: 150 million instructions. Note that, in general, staggering should be in the order of 100$\mu$s or 1ms, as shown in \cite{SergiDFT}, which would require staggering between a few hundreds of thousands or a few millions of instructions in most multicores. Nevertheless, SafeSoftDR is agnostic to the actual threshold provided and, if set too short, it could be the case that negative staggering values were observed, meaning that the trail process caught up with the head one so that diversity could be lost.

Figure~\ref{fig:result} shows the staggering observed during some consecutive executions of the SafeSoftDR monitor (\texttt{protect\_default} function). We observe that, in the third interval, the staggering drops down to $\approx$84M instructions. Then, the monitor stalls the trail process. For some intervals, the staggering remains below the threshold so that trail process remains stalled (staggering not plotted). Eventually, the staggering is sufficient (5$^{th}$ line in the figure), and the trail process resumed. In the next interval, the staggering falls below the threshold again, and the trail process is stalled. This sequence of events repeats several times until, after the last time the trail process is resumed in the figure, the staggering starts growing and stabilizing above 200 million of instructions. Hence, during that period, no trail process stalling occurs.
Overall, our example illustrates how SafeSoftDR monitors the progress of both processes created and enforces staggering. 

\textbf{Future prospects}. During the next months, our goal is to complete the porting of SafeSoftDR to an appropriate RISC-V platform. In particular, we are performing this porting to the H2020 SELENE RISC-V platform~\cite{SELENEgit}, which has been developed to support high-performance safety-related applications. Once such porting is complete, SafeSoftDR will be offered as an open-source component, and it will be compared against those solutions relying on hardware support, such as SafeDE and SafeDM, which are either already integrated into the SELENE platform (SafeDE) or will be in the forthcoming months (SafeDM). 

\section{Summary}

Enabling diverse redundancy on COTS processors becomes increasingly important to meet the safety requirements of high-integrity applications on platforms delivering enough performance. While we provided a proof of concept of a feasible software-only solution recently, it was just prototyped for a handcrafted example. In this paper, we present a standard interface offered in the form of a user-friendly library. It has already been shown to work on Linux on Intel and ARM-based platforms, and it is currently being integrated on RISC-V. We plan to offer this library as an open-source component in the forthcoming months once its integration on RISC-V and validation completes.

\section*{Acknowledgments}
This work is part of the project PCI2020-112010, funded by MCIN/AEI/10.13039/501100011033 and the European Union ``NextGenerationEU''/PRTR, and the European Union's Horizon 2020 Programme under project ECSEL Joint Undertaking (JU) under grant agreement No 877056.
This work has also been partially supported by the Spanish Ministry of Science and Innovation under grant PID2019-107255GB-C21 funded by MCIN/AEI/10.13039/501100011033.

\bibliographystyle{plain}  
\bibliography{biblio}

\end{document}